# Goldstone mode and pair-breaking excitations in atomic Fermi superfluids

Sascha Hoinka[1], Paul Dyke[1], Marcus G. Lingham[1], Jami J. Kinnunen[2], Georg M. Bruun[3] and Chris J. Vale[1*]

[1]*Centre for Quantum and Optical Sciences,Swinburne University of Technology, Melbourne 3122, Australia.*
[2]*COMP Centre of Excellence, Department of Applied Physics, Aalto University School of Science, FI 00076 Aalto, Finland.*
[3]*Institut for Fysik og Astronomi, Aarhus Universitet, 8000 Aarhus C, Denmark.*

*To whom correspondence should be addressed; E-mail: cvale@swin.edu.au

Spontaneous symmetry breaking is a central paradigm of elementary particle physics [1], magnetism [2], superfluidity [3] and superconductivity [4]. According to Goldstone's theorem, phase transitions that break continuous symmetries lead to the existence of gapless excitations in the long-wavelength limit [5]. These Goldstone modes generally dominate the low-energy excitations, showing that symmetry breaking has a profound impact on the physical properties of matter. Here, we present the first comprehensive study of the elementary excitations in a homogeneous strongly interacting Fermi gas through the crossover from a Bardeen-Cooper-Schrieffer (BCS) superfluid to a Bose-Einstein condensate (BEC) of molecules using two-photon Bragg spectroscopy. The spectra exhibit a discrete Goldstone mode, associated with the broken symmetry superfluid phase, as well as pair breaking single-particle excitations. Our techniques yield a direct determination of the superfluid pairing gap and speed of sound in close agreement with a strong-coupling theory.

When a Hamiltonian is invariant with respect to a continuous symmetry, but the ground state is not, a massless bosonic mode appears in the spectrum of allowed excitations [5]. At temperatures low enough for quantum effects to become prominent, dynamical behaviours, like superconductivity and superfluidity are only possible due to the low-energy excitation spectrum. Superfluid and superconducting states break gauge invariance and the resultant Goldstone mode is an oscillation of the phase of the corresponding order parameter giving rise to a collective motion of particles that is distinct from single-particle excitations. In superconductors, the Coulomb interaction lifts the collective mode up to the frequency of the classical plasma oscillation [4], present in the normal phase, such that the Goldstone mode is generally imperceptible [6]. In neutral superfluids however, the Goldstone mode takes the form of a gapless phonon [7] and provides a dramatic signature of macroscopic order.

Ultracold gases of atomic fermions have enabled the creation and study of high transition temperature superfluids in the smooth crossover from the BCS to BEC regimes [8]. Both first [9–11] and second [10] sound propagation have been observed in inhomogeneous Fermi gases, yet the basic elementary excitation spectrum has not been measured. Here, we present the first comprehensive measurements of the low-energy excitation spectrum in a homogeneous Fermi superfluid throughout the whole BCS-BEC crossover. The spectra exhibit both a dominant Goldstone mode, or Bogoliubov-Anderson (BA) phonon, and a single-particle continuum. Our study reveals how the energy and spectral weight of these excitations evolve as a function of the interaction strength. A theoretical model based on the quasiparticle random-phase approximation (QRPA) provides a very good quantitative description of the data.

The starting point for our experiments is a harmonically trapped gas of fermionic $^6$Li atoms in a balanced mixture of the lowest two hyperfine states with tunable $s$-wave interactions near a broad Feshbach resonance (see Supplementary Information). Atoms are cooled to temperatures below the superfluid transition temperature, $T_c$, across a wide range of the BCS-BEC crossover [12]. We measure the density-density response of these gases using two-photon Bragg spectroscopy [13–15]. This is the cold-atom analogue of inelastic neutron scattering, which has enabled characterisation of the dispersion relation, roton minimum, and condensate fraction in superfluid helium [3]; and inelastic X-ray scattering, used to measure electronic excitations in strongly-correlated materials [16]. Using tightly focussed Bragg lasers that intersect in the centre of a trapped atom cloud, we scatter atoms from a region of near homogeneous density (Fig. 1a). Bragg scattering involves the absorption of a photon with energy $\hbar\omega_a$, where $\hbar$ is Planck's constant, and wave vector $\mathbf{k}_a$ from one laser beam, and stimulated emission of a photon with energy $\hbar\omega_b$ and wave vector $\mathbf{k}_b$ into a second beam, thereby transferring energy $\hbar\omega = \hbar(\omega_a - \omega_b)$ and momentum $\hbar\mathbf{k} = \hbar(\mathbf{k}_a - \mathbf{k}_b)$. Within linear response the total momentum imparted to the cloud, $P_x$, is proportional to the dynamic susceptibility or density-density response function, $\mathrm{Im}\mathcal{D}(k,\omega)$ (see Supplementary Information).

Bragg spectra are obtained by applying a 1.2 ms Bragg pulse and measuring $P_x$ as a function of the Bragg frequency $\omega$. A Bragg spectrum provides full information on the energy and spectral weight of particle-conserving excitations at a particular momentum $\hbar\mathbf{k}$ [17]. After applying the Bragg pulse, atoms are released from the trap and we determine the relative centre-of-mass displacement, $\Delta X \propto P_x$, of the Bragg scattered volume, with respect to the unperturbed atoms, after 2 ms time of flight (TOF) [15]. In the measurements that follow, the mean atomic density in the Bragg volume, $\bar{n}$, lies in the range $0.90 \leq \bar{n}/n_0 \leq 0.95$, where $n_0$ is the peak density in the trap centre (see Supplementary Information).

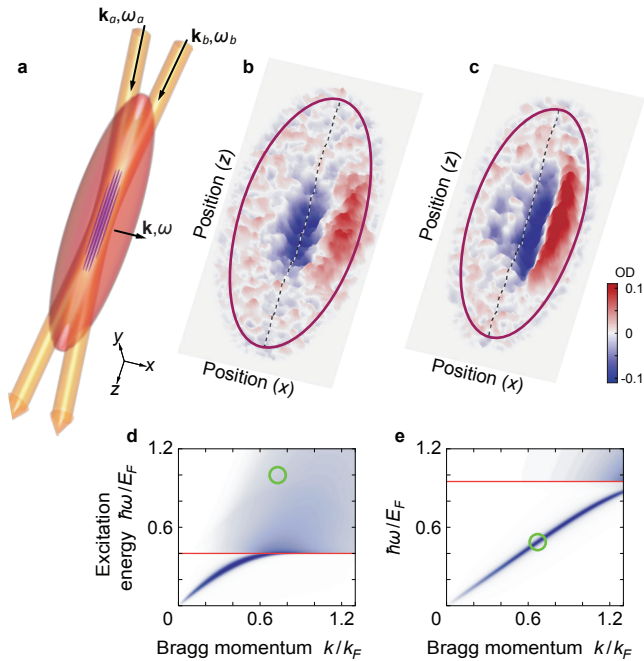

FIG. 1. **Focussed beam Bragg scattering. a**, Two far-detuned Bragg lasers, focussed to 20 $\mu$m $1/e^2$ radii, intersect at an angle of 12.9° in the centre of a harmonically trapped Fermi gas. **b,c**, Difference between images of atom clouds with and without Bragg scattering of (**b**) single particles above the pair breaking continuum at $1/(k_F a) = -0.6$ in the BCS regime, and, (**c**) the BA phonon at unitarity. Purple ellipses indicate the size of the expanded clouds and the dashed lines intersect the cloud centre. Each image optical density (OD) image is 400 $\mu$m by 180 $\mu$m. The Bragg laser intensities used in **b** were 2.4 times higher than those in **c** due to the weaker response of single particles compared to the collective BA mode. **d,e**, Calculated excitation spectra for $1/(k_F a) =$ -0.6, 0.0, respectively. Green circles mark the $(\omega, k)$ coordinates of the Bragg excitation used in **b** and **c** and the red lines indicate the threshold for single-particle excitations ($2\Delta$).

Fig. 1b shows the difference between two TOF images of atom clouds, with and without Bragg scattering, on the BCS side of the Feshbach resonance for an interaction strength of $1/(k_F a) = -0.6$ (875 G), where $k_F = (3\pi^2 \bar{n})^{1/3}$ is the Fermi wave vector and $a$ is the $s$-wave scattering length. Bragg scattering removes atoms from the centre of the cloud, where the lasers overlap, and displaces them along $x$. Here, $\omega$ lies in the single-particle continuum, indicated by the green circle on the excitation spectrum (Fig. 1d), calculated for $1/(k_F a) = -0.6$ using the QRPA theory (see Supplementary Information). The Bragg frequency is normalised by the Fermi energy, which is set by the mean density $E_F = k_B T_F = \hbar^2 k_F^2/(2m)$,

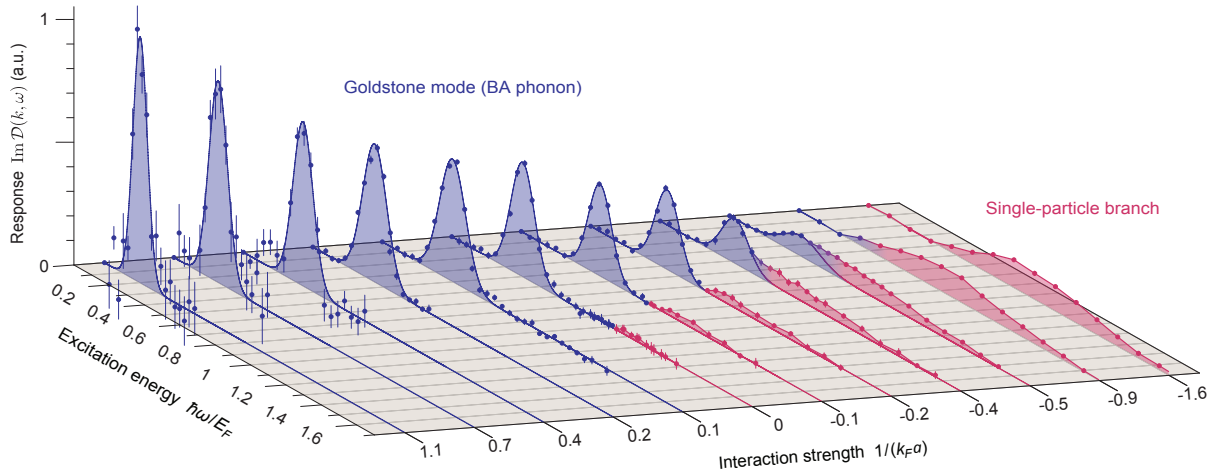

FIG. 2. **Bragg spectra throughout the BCS-BEC crossover**. Blue and red points are experimental data, solid blue lines are Gaussian fits to the BA mode peak (shaded blue) and the red shaded region indicates the single-particle excitation branch. The Bragg laser intensities are varied as the interactions are tuned from the BCS to BEC regime to maximise the signal-to-noise while remaining in the linear response regime (see Supplementary Information). $\text{Im}\mathcal{D}(k,\omega)$ plotted above has been scaled by the ratio of the relative two-photon transition probabilities so that all spectra can be displayed on the same scale.

where $k_B$ is Boltzmann's constant and $m$ is the atomic mass.

Fig. 1c shows a difference image taken with $\omega$ tuned to the peak of the BA phonon mode (Fig. 1e) at unitarity $1/(k_F a) = 0.0$. Comparing Fig. 1b with 1c reveals a qualitative difference between single-particle and phonon (collective) excitations. Phonons are a density modulation (sound wave) with the same periodicity as the Bragg lattice, excited when the lattice velocity ($\omega/k$) equals the sound velocity $c_s$. As the phonon propagates along $x$ and the cloud expands, the density continuously decreases and so too does the local speed of sound (gradient of the dispersion). In order to conserve energy and momentum, the phonon decays into lower energy modes, whose wave vectors need not be parallel, in a manner reminiscent of Beliaev damping [18]. While the density dependent decay dynamics may be quite complex, the qualitative result is a long wavelength density modulation propagating along $x$ (Fig. 1c). In contrast to the case of single-particle scattering, the density minimum no longer remains fixed at the location of the Bragg lasers.

We have measured a series of Bragg spectra at $k \sim k_F/2$, within the linear response regime, for a range of interaction strengths, plotted in Fig. 2 (see Supplementary Information). At unitarity the local temperature within the Bragg volume is $T/T_F = 0.09(1)$, determined by fitting to the known equation of state [19]. While precise thermometry away from unitarity remains a challenge, tuning to the BCS (BEC) regime by scanning the magnetic field adiabatically around the Feshbach resonance will decrease (increase) $T/T_F$ such that we expect to remain below $T_c$ for all spectra, with the possible exception of $1/(k_F a) = -1.6$ [12]. However, for $1/(k_F a) = -0.9$ and $-1.6$, (rightmost traces in Fig. 2) we observe only single-particle excitations. This is because the pair correlation length, $\xi_{pair}$ [20], is significantly greater than $1/k$ and the phonon mode merges with the single-particle continuum well below $k_F/2$ [21, 22]. Closer to the Feshbach resonance, the Goldstone or BA phonon mode (shaded blue) first appears around $1/(k_F a) \lesssim -0.5$, which corresponds to $\xi_{pair} \approx 1/k$ [20], where it becomes well defined. At unitarity, the BA mode dominates the spectra and the single-particle branch is increasingly suppressed. In the BEC regime single-particle excitations fall below our measurement sensitivity due to the reduced spin-susceptibility [23] as fermion pairs become more tightly bound.

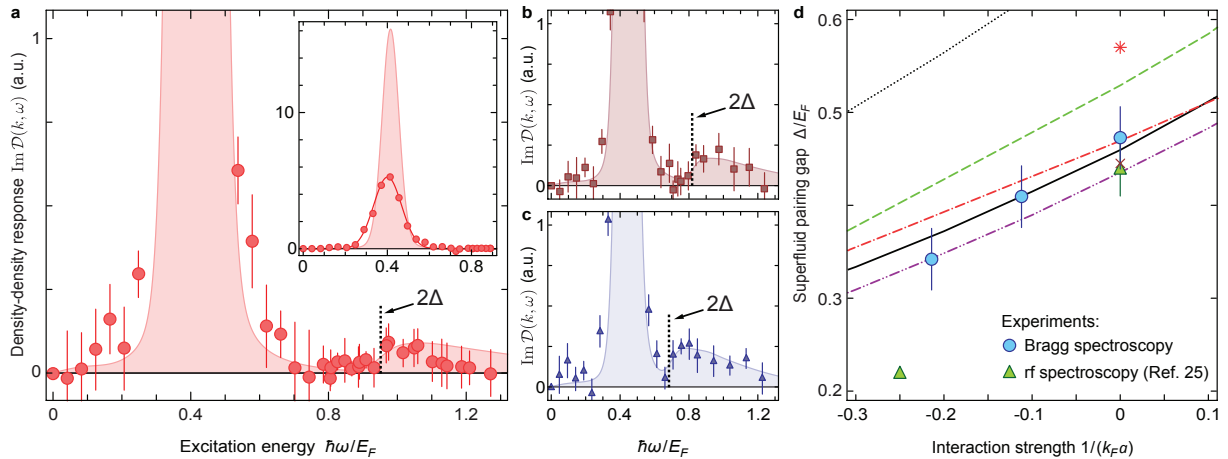

FIG. 3. **Bragg spectra and pairing gap near unitarity**. **a**, Experimental spectrum at unitarity (filled data points) and QRPA theory (shaded). Vertical dashed lines indicate the onset of the single-particle branch for $\hbar\omega \geq 2\Delta$. Inset: Comparison of experimental and theoretical BA mode peaks. **b,c**, Zoomed view of Bragg spectra at $1/(k_F a) = -0.11$ and $1/(k_F a) = -0.21$. **d**, Pairing gap $\Delta$ determined from Bragg spectra (blue points) along with previous rf measurements (green triangles) [25] and theoretical predictions: BCS (black dotted line), T-matrix (green dashed) [26], Luttinger-Ward (black solid) [12], extended mean-field (red dash-dotted) [28], Monte Carlo (brown cross) [27], operator product expansion (red asterisk) [29], and extended T-matrix (purple dash-dot-dotted) [30].

In the range $-0.5 \leq 1/(k_F a) \leq 0$, both phonon and single-particle excitations are visible in the individual spectra. Furthermore, for $-0.2 \leq 1/(k_F a) \leq 0$, these two branches separate from each other, enabling direct read-off of the superfluid pairing gap $\Delta$. Fig. 3a shows a zoomed view of the Bragg spectrum at unitarity. Filled points are experimental data and the shaded curve is the QRPA theory including the Fourier width of the Bragg pulse. Since a minimum energy of $\hbar\omega = 2\Delta$ is required to break a pair and produce two free atoms, we associate the sharp onset of single-particle excitations with $2\Delta$ (dashed vertical line). At unitarity we find $\Delta/E_F = 0.47 \pm 0.03$. The pairing gap has previously been measured using momentum [24] and spatially resolved [25] radio frequency (rf) spectroscopy. Our localised Bragg measurements are consistent with previous rf data [25], yet provide the pairing gap directly, free of final state effects, Hartree energy shifts or density inhomogeneities. Fig. 3b and 3c show spectra for $1/(k_F a) = -0.11$ and $1/(k_F a) = -0.21$, respectively. A plot of $\Delta/E_F$ for these three spectra is provided in Fig. 3d along with different theoretical calculations [12, 26–30].

The experimentally determined pairing gap can serve as an input parameter for the QRPA calculation (see Supplementary Information). This leads to striking agreement between theory and experiment over the full excitation spectrum (shaded curves, Fig. 3), particularly concerning the frequency of the BA mode. However, when the theory is scaled to match the amplitude of the experimentally measured single-particle branch, the calculated BA mode peak is consistently more than twice as high and approximately two-thirds as wide as the measurement (inset, Fig. 3a). Damping mechanisms [31], not included in the theory, may be responsible for this discrepancy.

As $k \to 0$, the BA mode displays linear dispersion with a gradient set by the sound velocity $c_s$. The centre frequency of the BA mode $\omega_{BA}$ is found from a Gaussian fit to the BA peak (solid blue lines in Fig. 2) providing a measure of $c_s/v_F = \omega_{BA}/(kv_F)$, where $v_F = \hbar k_F/m$ is the Fermi velocity. This is plotted in Fig. 4 (blue circles) for interactions where we obtain a reliable fit. In the momentum range used here, $0.4 \lesssim k/k_F \lesssim 0.6$ (see Supplementary Information), the dispersion remains close to linear near unitarity, but becomes concave (convex) in the BCS (BEC) regime [22, 31]. Based on the QRPA calculation we can estimate the curvature for finite $k/k_F$ and correct for this to obtain $c_s/v_F$ in the $k \to 0$ limit (see Supplementary Information). The corrected data (orange diamonds in Fig. 4) agree well with theoretical calculations of $c_s/v_F$ throughout the BCS-BEC crossover [12, 28, 30, 32]. Our experiments are the first to study sound propagation in a homogeneous gas and show good qualitative agreement with previous trap-averaged measurements using inhomogeneous clouds [9, 11]. At unitarity, $c_s/v_F$ is set by $\sqrt{\xi/3}$ as $T \to 0$, where $\xi$ is the Bertsch parameter (ratio of the internal energy of a resonantly interacting Fermi gas to that of an ideal Fermi gas). Our value of $c_s/v_F = 0.36 \pm 0.02$ yields $\xi = 0.39 \pm 0.04$ consistent with thermodynamic measurements [19].

We have performed the first comprehensive study of the elementary excitations in a homogeneous Fermi gas throughout the BCS-BEC crossover. Focussed beam Bragg spectroscopy allows the unambiguous measurement of the Goldstone mode and pair breaking single-particle excitations, that directly yield the sound velocity and pairing gap near unitarity. Our data reveal the canonical excitation spectra of Fermi superfluids with tunable $s$-wave pairing, that are difficult to access in other systems such as superconductors, thereby establishing quantitative benchmarks for many-body theories of strongly correlated fermions. This work paves the way for studies of the elementary excitations in a range of systems, including lower-dimensional and spin-orbit coupled Fermi gases, and provides a means to characterise pseudogap phenomena in these systems.

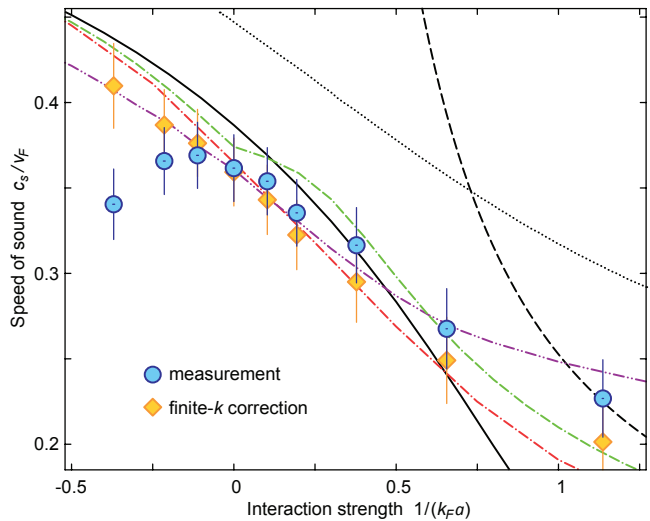

FIG. 4. **Speed of sound across the BCS-BEC crossover**. Solid blue points are taken from the raw BA mode frequency and orange diamonds show corrected estimates of $c_s$ as $k \to 0$ based on the calculated curvature of the BA mode dispersion at nonzero $k/k_F$ (see Supplementary Information). Also shown are theoretical predictions for $c_s/v_F$: BCS theory (black dotted line), Beliaev (black dashed) [18], density functional (green dash-dash-dotted) [32], Luttinger-Ward (black solid) [12], extended mean-field (red dash-dotted) [28], and extended T-matrix (purple dash-dot-dotted) [30].

**Acknowledgements:** We thank W. Zwerger, G. Strinati, L. Salasnich and Y. Ohashi for sharing their data and comments on the manuscript and P. Hannaford for fruitful discussions. This work was supported by ARC Discovery Project DP130101807.

**Author contributions:** S. H., P. D., M. G. L and C. J. V conducted the experimental work and data analysis. J. J. K and G. M. B. performed the theoretical calculations. All authors contributed to the manuscript preparation.

# Goldstone mode and pair-breaking excitations in atomic Fermi superfluids

Sascha Hoinka[1], Paul Dyke[1], Marcus G. Lingham[1], Jami J. Kinnunen[2], Georg M. Bruun[3] and Chris J. Vale[1]*

[1]Centre for Quantum and Optical Sciences,
Swinburne University of Technology, Melbourne 3122, Australia

[2]COMP Centre of Excellence, Department of Applied Physics,
Aalto University School of Science, FI 00076 Aalto, Finland

[3]Institut for Fysik og Astronomi,
Aarhus Universitet, 8000 Aarhus C, Denmark

*To whom correspondence should be addressed; E-mail: cvale@swin.edu.au

# Supplementary Material

## 1 Cloud preparation and Bragg spectroscopy

In our experiments, fermionic $^6$Li atoms from a Zeeman slowed atomic beam are cooled in a magneto-optical trap and loaded into a 100 W, 1075 nm single beam optical dipole trap. An external magnetic field is tuned to 832 G and a radio frequency (rf) field is applied to produce a balanced mixture of the $|F=1/2, m_F = \pm 1/2\rangle$ spin states, labelled $|\uparrow, \downarrow\rangle$, at the pole of the broad *s*-wave Feshbach resonance [1]. The gas is then evaporatively cooled by lowering the power of the dipole trap over a few seconds. Subsequently, the atoms are transferred into a deep, highly harmonic hybrid optical (1064 nm) and magnetic trap with confinement frequencies at unitarity of $(\omega_x, \omega_y, \omega_z) = 2\pi \times (109, 101, 24.5)$ s$^{-1}$. The axial (*z*) confinement is provided by the residual field curvature from the Feshbach magnetic coils. We obtain typically $N/2 \approx 3 \times 10^5$ atoms per spin state and a final temperature of 0.09(1) of the Fermi temperature $T_F$ at unitarity. The temperature is determined by fitting the known equation of state for the pressure of a unitary Fermi gas [2] to the line densities of trapped atom clouds [3].

Bragg spectra are obtained for a homogeneous Fermi gas at various interaction strengths, i.e different magnetic fields, throughout the BCS-BEC crossover via local Bragg spectroscopy. For this, two intensity stabilised laser beams are focussed to $1/e^2$ radii of 20 µm to illuminate a volume with near uniform density in the central region of the trapped atom cloud. The angle $2\theta$ between the beams is set to $12.9° \pm 0.2°$, giving a Bragg wave vector of $|\mathbf{k}| = k = (4\pi/\lambda)\sin\theta = 2.12(4)$ µm$^{-1}$ for a wavelength $\lambda = 671$ nm. The alignment of the Bragg laser beams is controlled to micrometer precision using piezo-activated mirror mounts. The Bragg lasers generate a perturbation of the atomic density, where, within the Bragg volume, a spatially modulated interference pattern with $\lambda/(2\sin\theta) \simeq 3$ µm lattice spacing moves in $x$-direction at velocity $\omega/k$ (Fig. 1a main text). To obtain a Bragg spectrum, the frequency difference $\omega$ between the two laser beams is scanned typically over the range 0 to $\pm 2\pi \times 15$ kHz. Both positive and negative Bragg frequencies are used and the results are averaged for improved signal to noise. After pulsing the Bragg lasers on for 1.2 ms, the atoms are immediately released from the trap. Such a pulse duration along with a Gaussian pulse envelope yields a FWHM spectral resolution of $\sim 1$ kHz.

## 2 Linear response for Bragg spectroscopy

The coupling between the Bragg lasers and atomic states can be expressed by the two-photon Rabi frequency $\Omega_{Br}(\mathbf{r}) = \Gamma^2 \sqrt{I_a(\mathbf{r})I_b(\mathbf{r})}/(4\Delta_\sigma I_{sat})$, where $\Gamma$ is the natural line width of the Bragg transition, $I_{sat}$ is the saturation intensity, $I_{a,b}(\mathbf{r})$ are the laser intensities with Gaussian spatial profiles, and $\Delta_\sigma$ is the laser detuning relative to the atomic transitions from the spin state $|\sigma\rangle$. The Bragg lasers are detuned very far from the atomic transitions (between 100 GHz and 1 THz) so that both spin states couple equally ($\Delta_\uparrow \equiv \Delta_\downarrow$) and we probe the density-density response [4]. For a weak density perturbation (neglecting the spatial dependence of $\Omega_{Br}$), the operator describing the Bragg light-matter coupling is [5]

$$H_{Br} = \frac{\hbar \Omega_{Br}}{2}(e^{-i\omega t}\hat{\rho}_{\mathbf{k}}^\dagger + e^{i\omega t}\hat{\rho}_{\mathbf{k}}), \tag{1}$$

where $\hat{\rho}_{\mathbf{k}} = \sum_{\mathbf{p},\sigma} \hat{a}_{\mathbf{p}-\mathbf{k},\sigma}^\dagger \hat{a}_{\mathbf{p},\sigma}$ is the Fourier transform of the density operator. Here, $\hat{a}_{\mathbf{p},\sigma}$ removes an atom with momentum $\mathbf{p}$ and spin $\sigma$. The rate of momentum transfer to the cloud due to the Bragg pulse within linear response is given by [5]

$$\frac{d\mathbf{P}}{dt} = -\frac{\hbar \mathbf{k} \Omega_{Br}^2}{2} \operatorname{Im}\mathcal{D}(\mathbf{k},\omega), \tag{2}$$

where $\mathcal{D}(\mathbf{k},\omega)$ is the Fourier transform of the retarded density-density correlation function $\mathcal{D}(\mathbf{r}-\mathbf{r}',t-t') = -i\theta(t-t')\langle[\hat{n}(\mathbf{r},t),\hat{n}(\mathbf{r}',t')]\rangle$, which contains information on both single-particle and collective excitations. Here, $\hat{n}(\mathbf{r})$ is the total density operator and $\langle\ldots\rangle$ denotes the thermal average. Thus, Bragg scattering probes the imaginary part of the dynamic susceptibility or density-density response function.

## 3 Obtaining Bragg spectra

The response to a Bragg pulse in our experiments is quantified by the momentum $\mathbf{P}$ imparted to the cloud. The momentum is determined by measuring the centre-of-mass cloud displacement $\Delta X$ which relates to the $x$-component of $\mathbf{P}$ through the centre-of-mass velocity $\Delta X/\Delta t \propto P_x$, where $\Delta t$ is the time of flight (TOF) of a freely expanding cloud [6]. Following the Bragg pulse, we switch off the trap and take an absorption image of the cloud after 2 ms TOF. An example image is given in Supp. Fig. 1a for $\omega = 2\pi \times 4$ kHz and $1/(k_F a) = 0$ corresponding to the peak of the phonon mode.

The homogeneous response is determined in the following way: we evaluate $\Delta X$ by integrating the cloud image vertically over a narrow strip (red shaded area in Supp. Fig. 1) which contains the Bragg signal and computing the first moment of the resultant line profile (Supp. Fig. 1a, lower panel). As the Bragg lasers are focussed into the centre of the cloud (Fig. 1a main text), the response originates from a region of near uniform density. We then subtract the centre of mass of the unperturbed cloud wings (green shaded areas in Supp. Fig. 1) where little or no Bragg scattering can be detected (Supp. Fig. 1b). We find that fitting a Gaussian function to the cloud wings provides the most robust reference for the centre of mass as the fit is less sensitive to noise or any asymmetries in the cloud profile. The size of the wing region is chosen to be as large as possible to maximise the signal to noise. The extent of the central regions of interest (red shaded area in Supp. Fig. 1) along $z$ are chosen such that the mean density in the Bragg volume stays within 10% of the peak density in the trap centre $n_0$. To further improve the signal to noise, many data points ($> 20$) are taken for each $\omega$, and measurements using both positive and negative $\omega$ are averaged to obtain a Bragg spectrum.

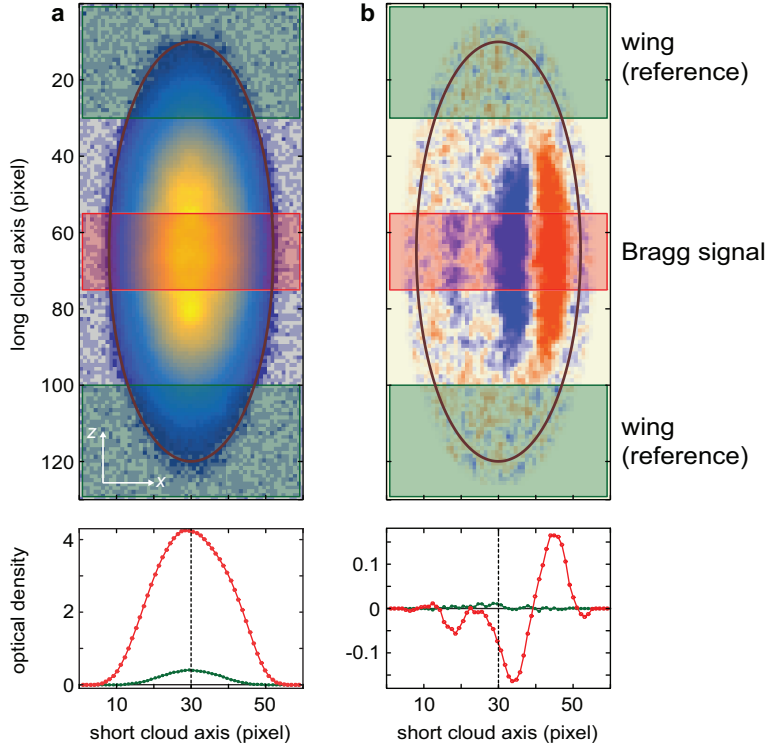

Supplementary Figure 1: **2D and 1D optical density profiles of a Bragg scattered cloud. a**, Absorption image (false color) of a unitarity cloud after a Bragg pulse at $\omega = 2\pi \times 4$ kHz and 2 ms TOF. The red (green) shaded area indicates the region of interest for computing the centre of mass or first moment with (without) Bragg scattering. **b**, Difference of image in **a**, and a cloud without Bragg scattering, showing atoms displaced by the Bragg pulse. Lower panels are the integrated line profiles for the red and green shaded regions of interest. The spatial units are given in camera pixels, where 1 pixel has an effective area of $2.84 \times 2.84$ μm$^2$.

## 4 Determination of the mean density in the Bragg volume

An important quantity for our measurements is the mean density $\bar{n}$ inside the Bragg volume, which is given by the convolution of the true (3D) density distribution $n(\mathbf{r})$ of the trapped cloud, and, the intensity product of the Bragg laser beams with their intersecting Gaussian spatial profiles, which sets the scattering rate, Eq. 2. We

therefore define

$$\bar{n} = \frac{\int n(\mathbf{r})\,\Omega_{Br}^2(\mathbf{r})\,\mathrm{d}^3 r}{\int \Omega_{Br}^2(\mathbf{r})\,\mathrm{d}^3 r}, \tag{3}$$

where the integration is carried out over the entire *xy*-plane and extends over the vertical region of interest in the *z* direction (red shaded area in Supp. Fig. 1). This mean density sets the Fermi energy $E_F = \frac{\hbar^2}{2m}(3\pi^2\bar{n})^{2/3}$, Fermi wave vector $k_F = (3\pi^2\bar{n})^{1/3}$ and Fermi velocity $v_F = \frac{\hbar k_F}{m}$, where *m* is the atomic mass.

We obtain the 3D density $n(\mathbf{r})$ from the inverse Abel transform of absorption images of trapped clouds, in the absence of Bragg scattering, taking into account the elliptic symmetry of the density distribution [2]. To image these clouds with high optical density ($OD \sim 5$), we use a short imaging pulse (1 μs) and employ the procedure described in Ref. [7] to find a correction factor to calibrate the absolute scale of the density. The inverse Abel method we use is based on Fourier decomposition where the radial density distribution is expanded in a Fourier series [8]. As this method does not rely on direct differentiation, it is possible to reconstruct the central density ($\mathbf{r} \to 0$) of the atomic cloud without singularities. It also allows filtering of high spatial frequency noise from the image. Using trial functions, we have verified that systematic errors introduced in obtaining $n(\mathbf{r})$ by this method are below 1%. With this and the density calibration we estimate the error of $\bar{n}$ to be around 5 %.

## 5 Experimental parameters

The tables below summarise all relevant parameters for the measurements presented in the paper. For the results in Fig. 1 of the main text, the parameters are given in Supp. Table 1, and for Fig. 2 to 4, they are listed in Supp. Table 2. The uncertainty in the measured mean densities $\bar{n}$ of the Bragg volume is estimated to be 5% and just over 3% for the Fermi energy. The main contribution to the uncertainty in the Bragg wave vector ($k/k_F$) comes from the measured angle between the two Bragg laser beams (12.9° ± 0.2°), whereas for the gap ($\Delta/E_F$), the dominant contribution originates from the Fourier limited spectral resolution of the Bragg pulse. Note, where uncertainties are quoted in a column in Supp. Tables 1 and 2, the last value in brackets also applies to the remaining numbers below in that column.

| $1/(k_F a)$ | $B$ (G) | $\Omega_{Br}$ (Hz) | $\bar{n}$ (μm$^{-3}$) | $E_F$ (kHz) | $k/k_F$ |
|---|---|---|---|---|---|
| 0.0 | 832.2 | 240 | 1.06 | 8.4 | 0.67(2) |
| -0.58 | 875 | 580 | 0.81 | 7.0 | 0.73 |

Supplementary Table 1: Experimentally obtained values for the interaction strength $1/(k_F a)$, Rabi frequency $\Omega_{Br}$, mean density $\bar{n}$, Fermi energy $E_F$, and Bragg wave vector $k/k_F$ at given magnetic field $B$.

As seen in the Supp. Tables 1 and 2, we have used different values for the Rabi frequency across the BCS-BEC crossover. This is because the density-density response grows significantly from the BCS to the BEC regime. Suitable combinations of Bragg laser intensity and detuning were therefore chosen to remain in the linear response regime [9] for all interaction strengths while maximising signal to noise. In Fig. 2 of the main text, the measured centre-of-mass displacements in the Bragg spectra obtained for $\Omega_{Br}$ = 90 Hz and 700 Hz were scaled by the square of the ratio of the Rabi frequencies (i.e. $(90/270)^2$ and $(700/270)^2$, respectively) to match the spectra at $\Omega_{Br}$ = 270 Hz. As a further check of the linearity of our measurements, additional Bragg spectra were obtained at $1/(k_F a) = -0.49$ and 0.19 using lower Rabi frequencies of 270 Hz and 90 Hz, respectively. When scaled by the square of the ratio of two-photon Rabi frequencies, the height and shape of both sets of spectra are the same within our experimental error bars.

| $1/(k_F a)$ | $B$ (G) | $\Omega_{Br}$ (Hz) | $\bar{n}$ (μm$^{-3}$) | $E_F$ (kHz) | $k/k_F$ | $(c_s/v_F)_{(meas.)}$ | $(c_s/v_F)_{(corr.)}$ | $\Delta/E_F$ |
|---|---|---|---|---|---|---|---|---|
| 1.14 | 750 | 90 | 3.57 | 18.8 | 0.44(2) | 0.23(2) | 0.20(3) | |
| 0.66 | 780 | 90 | 3.07 | 17.0 | 0.47 | 0.27 | 0.25(3) | |
| 0.38 | 803 | 90 | 2.08 | 13.1 | 0.53 | 0.32 | 0.30(2) | |
| 0.19 | 816 | 270 | 2.23 | 13.7 | 0.52 | 0.34 | 0.32 | |
| 0.10 | 824 | 270 | 1.78 | 11.8 | 0.56 | 0.35 | 0.34 | |
| 0.0 | 832.2 | 270 | 1.85 | 12.1 | 0.55 | 0.36 | 0.36 | 0.47(3) |
| -0.11 | 841 | 270 | 1.43 | 10.2 | 0.60 | 0.37 | 0.38 | 0.41 |
| -0.21 | 850 | 270 | 1.52 | 10.6 | 0.59 | 0.37 | 0.39 | 0.34 |
| -0.37 | 863 | 270 | 1.31 | 9.6 | 0.62 | 0.34 | 0.41 | |
| -0.49 | 875 | 700 | 1.36 | 9.9 | 0.61 | | | |
| -0.93 | 920 | 700 | 1.12 | 8.7 | 0.65 | | | |
| -1.58 | 1032 | 700 | 1.09 | 8.5 | 0.66 | | | |

Supplementary Table 2: Experimentally obtained values for the interaction strength $1/(k_F a)$, Rabi frequency $\Omega_{Br}$, mean density $\bar{n}$, Fermi energy $E_F$, Bragg wave vector $k/k_F$, measured speed of sound $c_{s(meas.)}$, corrected speed of sound $c_{s(corr.)}$, and pairing gap $\Delta$ at given magnetic field $B$.

# 6 $f$-sum rule

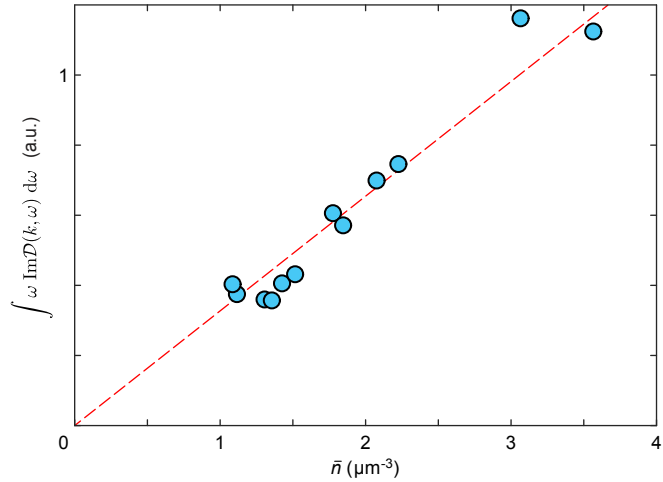

Supplementary Figure 2: $f$**-sum rule.** First energy-weighted moment of all Bragg spectra plotted Fig. 2 of the main text vs. the mean density within the Bragg scattered volume (blue points). Red line is a straight line fit to the data showing that the moment scales linearly with the measured density in accordance with the $f$-sum rule.

A further check that our Bragg spectra have been obtained in the linear response regime is possible through the application of sum rules. An important sum rule, directly connected to particle number conservation, is the

*f*-sum rule for the first energy-weighted moment of the dynamic structure factor $S(\mathbf{k},\omega)$ given by [10]

$$\hbar^2 \int_{-\infty}^{+\infty} \omega S(\mathbf{k},\omega) d\omega = N \frac{\hbar^2 k^2}{2m}, \qquad (4)$$

where $N$ is the total number of particles and $\frac{\hbar^2 k^2}{2m} = E_r$ is the recoil energy. Our measurements probe the imaginary part of the density-density response function $\operatorname{Im}\mathcal{D}(\mathbf{k},\omega)$ which is related to $S(\mathbf{k},\omega)$ via the fluctuation dissipation theorem:

$$\operatorname{Im}\mathcal{D}(\mathbf{k},\omega) = -\pi\left[S(\mathbf{k},\omega) - S(\mathbf{k},-\omega)\right]. \qquad (5)$$

It then follows that the *f*-sum rule for $\operatorname{Im}\mathcal{D}(\mathbf{k},\omega)$ is [10]

$$\hbar^2 \int_{-\infty}^{+\infty} \omega \operatorname{Im}\mathcal{D}(\mathbf{k},\omega) d\omega = 2\pi N E_r. \qquad (6)$$

We have evaluated the first energy-weighted moments of all spectra shown in Fig. 2 of the main text and the results are plotted in Supp. Fig. 2 as a function of the mean density $\bar{n}(\propto N)$ given in the above section. Low density data correspond to spectra taken in the BCS regime and the highest density data are from the BEC regime. It is clear from the figure that the moment follows the predicted linear scaling with mean density indicating that the data are obtained within the linear response regime. It also shows that our measurement covers the full frequency range where there is significant spectral weight, since the *f*-sum rule is exhausted. Finally, as the recoil energy and Bragg volume are fixed for all spectra, this verifies that the rescaling by the relative two-photon Rabi frequencies (described in the above section) preserves linearity.

## 7 Theory: quasiparticle random-phase approximation

We use a quasiparticle random-phase approximation (QRPA) combined with an extended mean-field theory to calculate the density-density response function $\mathcal{D}(\mathbf{k},\omega)$ in Eq. 2. The derivation of the QRPA theory is rather lengthy, therefore we discuss here only the physically important aspects of the theory. Details of this approach are described in Refs. [11, 12, 13, 14, 15, 16]. In QRPA, the response function has the form

$$\mathcal{D}(\mathbf{k},\omega) = \frac{\mathcal{D}_0(\mathbf{k},\omega)}{1 - \mathcal{T}\mathcal{L}(\mathbf{k},\omega)}, \qquad (7)$$

where $\mathcal{T} = \frac{4\pi\hbar^2 a}{m}$ is the two-body scattering matrix and $a$ the s-wave scattering length. The denominator in Eq. 7 is closely related to the mean-field superfluid gap equation (Eq. 8, see below). Indeed, one recovers the gap equation by equating the denominator to zero for $(\mathbf{k},\omega) = (0,0)$. And like the standard mean-field BCS-BEC crossover theory, the QRPA model is well-behaved across the BCS-BEC crossover. In a two-component superfluid system, the response function $\mathcal{D}(\mathbf{k},\omega)$ actually is a vector with four components, describing the response in the densities of the two spin components and in the pairing fields. Bragg spectroscopy probes the total density-density part of this response function, as seen from Eq. 2.

The bare response function $\mathcal{D}_0(\mathbf{k},\omega)$ in Eq. 7 gives the responses in terms of elementary particle-hole, particle-particle, and hole-hole excitations. For low momenta $(k < k_F)$, the minimum energy for creating such excitations in a zero-temperature superfluid is $2\Delta$, where $\Delta$ is the binding energy of the pairs (Cooper pairs in the BCS regime, bosonic dimers in the BEC regime). However, the form of the response in Eq. 7 shows that resonances can occur also when the denominator $1 - \mathcal{T}\mathcal{L}(\mathbf{k},\omega)$ vanishes. The $4 \times 4$ matrix $\mathcal{L}$ in Eq. 7 describes the coupling between fluctuations in the densities and the pairing fields, and a resonance describes a collective excitation. One of these excitations is the Goldstone mode originating in phase fluctuations of the $U(1)$ broken symmetry pairing field, which give rise to density fluctuations.

Note that to recover this gapless Goldstone mode in the long-wavelength limit (Eq. 7) imposes important constraints on the theory. In particular, the coupling matrix $\mathcal{L}(\mathbf{k},\omega)$ and the bare response function $\mathcal{D}_0(\mathbf{k},\omega)$ have to be calculated in a consistent way from the underlying static theory. One way to achieve this is to use the Kadanoff-Baym formalism for imaginary time Green's functions, but we will here use a simpler and more pragmatic approach.

# 8 Extended mean-field theory

Our strategy for calculating $\mathcal{D}_0(\mathbf{k},\omega)$ and $\mathcal{L}(\mathbf{k},\omega)$ is to use mean-field theory, which is extended to include fluctuations in the number densities in the spirit of Ref. [17]. These fluctuations are important away from the weakly interacting BCS regime, and the standard mean-field BCS-BEC crossover theory of Refs. [18, 19] is therefore insufficient for describing the experimental results. Within standard BCS theory, the superfluid gap equation reads

$$\Delta = \mathcal{T} \sum_k \frac{\Delta[1 - 2n(E_k)]}{2E_k}, \tag{8}$$

where $\Delta$ is the superfluid excitation gap, $E_k = \sqrt{\varepsilon_k^2 + \Delta^2}$ the quasiparticle energy, $\varepsilon_k = \frac{\hbar^2 k^2}{2m} - \mu$ the non-interacting single-particle energy minus the chemical potential $\mu$, and $n(E)$ is the Fermi-Dirac distribution. The number equation is

$$N = \sum_k \left(1 - \frac{\varepsilon_k}{E_k} + 2n(E_k)\frac{\varepsilon_k}{E_k}\right). \tag{9}$$

These two equations are solved self-consistently, yielding values for $\mu$ and $\Delta$. From this, one can then calculate the QRPA response function given by Eq. 7. However, BCS theory significantly overestimates the superfluid gap $\Delta$ as well as the critical temperature $T_c$ for strong coupling; we therefore need to improve the theory to explain the experimental data.

## 8.1 Fitting scheme and numerical parameters

There is no quantitatively accurate microscopic theory available in the whole BCS-BEC crossover, and we therefore adapt a more pragmatic approach. In the strongly interacting regime, we can measure the pairing gap $\Delta$ from the threshold of the pair-breaking excitation continuum. We then simply tune the chemical potential $\mu$ so that Eq. 8 yields the experimentally observed gap for a given temperature and coupling strength. The number equation, Eq. 9, is neglected at this point (see 8.2). Once we have determined the corresponding chemical potential, we use the QRPA theory to calculate the whole spectral response function.

While this obviously leads to a perfect overlap of the positions of the theoretical and experimental pair breaking thresholds, we emphasise that we get excellent agreement between theory and experiment for the *entire* spectral response as a function of frequency, for a wide range of interaction strengths. This is highly non-trivial for this strongly interacting system, and it illustrates the accuracy of our QRPA approach. In particular, the theory predicts a frequency of the BA mode consistent with the experimental result. This also enables us to interpolate the speed of sound even when the Bragg momentum is not in the linear regime.

In the BCS regime (interaction strengths $1/(k_\mathrm{F}a) = -0.37, -0.49$ in Supp. Table 3 below), the phonon branch and the pair breaking continuum merge so that the pairing gap can no longer be determined experimentally.

Also in the BEC regime ($1/(k_\text{F}a) = 0.66$ in Supp. Table 3), single-particle excitations are highly suppressed, hence $\Delta$ cannot be measured. In these regimes, we therefore determine the value of $\Delta$ by fitting the position of the QRPA phonon peak to the measured response maximum. In practice, we need to choose some value for $\Delta$, determine the corresponding chemical potential $\mu$ using the gap equation (8) (again neglecting the number equation), compare the QRPA spectrum with the experimental spectrum, and find a new guess for the pairing gap $\Delta$. We note that it is not possible to check the validity of our approach for these interactions strengths due to lack of experimental data. However, this is not a serious limitation as the most interesting region is the strong coupling unitarity regime, where the gap is experimentally available.

## 8.2 Discussion of the theoretical model

We would like to put our theoretical approach into a broader context and discuss how the calculated values of the chemical potential $\mu$ can be interpreted. Our approach corresponds to including beyond mean-field fluctuations in the number equation (9), whereas the gap equation (8) and the density-density response function (7) retain their simple form [17]. Including fluctuations in the number equation changes the chemical potential $\mu$ away from the value $\mu_\text{BCS}$ predicted by Eq. 9. The difference in these chemical potentials, $\mu - \mu_\text{BCS}$ can for zero temperature be interpreted as a self-energy shift $\Sigma_0$ at the underlying Fermi surface. Neglecting temperature effects, it follows from this interpretation that the magnitude of the self-energy $\Sigma_0$ can be determined by comparing the chemical potential $\mu$ obtained from the fitting procedure described above and the calculated chemical potential from Eq. 9. The results for the energy shift $\Sigma_0$, pairing gap $\Delta$, and chemical potential $\mu$ following the strategy explained above are shown in Supp. Table 3. They may be used for comparing with theoretical predictions for the self-energy shift at the Fermi surface.

| $1/(k_F a)$ | $\Sigma_0/E_F$ | $\mu/E_F$ | $\Delta/E_F$ |
|---|---|---|---|
| 0.66 | -0.88 | -0.31 | 0.72 |
| 0.0 | -0.39 | 0.41 | 0.47 |
| -0.11 | -0.38 | 0.47 | 0.41 |
| -0.21 | -0.38 | 0.51 | 0.34 |
| -0.37 | -0.38 | 0.56 | 0.25 |
| -0.49 | -0.27 | 0.66 | 0.25 |

Supplementary Table 3: Calculated values for the energy shift $\Sigma_0$, pairing gap $\Delta$, and chemical potential $\mu$ at given interaction strength $1/(k_F a)$.

Note that since the fluctuation contribution does not enter the gap equation, the QRPA respose function (7) must be calculated using the standard expressions obtained from BCS theory [11, 14, 15]. In this way, we ensure that we recover the gapless Goldstone mode, while the effects of strong coupling fluctuations are included through the renormalized values of $\mu$ and $\Delta$.

## 9 Correction to the speed of sound

The Bragg spectra presented in the main text were obtained using wave vectors $k_{meas.} \sim k_F/2$ (listed in Supp. Table 2). At these $k$, the phonon mode dispersion can show a departure from the linear behaviour found as $k \to 0$, particularly on the BCS side of the Feshbach resonance where back-bending occurs as the phonon mode approaches the single-particle continuum. In order to correct for this, we can use the calculated dispersions from the QRPA theory to estimate the speed of sound in the $k \to 0$ limit.

Experimentally, we determine the centre frequency of the phonon mode $\omega_{BA}$ at $k_{meas.}$ using a Gaussian fit to the phonon peak. This provides the quantity $c_{s(meas.)} = \omega_{BA}(k_{meas.})/k_{meas.}$. Based on calculated excitation spectra (as in Fig. 1d-e of the main text), for the interaction strengths given in Supp. Table 2, we evaluate $c_{s(th)}(k) = \omega_{BA}(k)/k$ for both $k \to k_{meas.}$ and as $k \to 0$. To obtain the corrected speed of sound $c_{s(corr.)}$ we simply multiply our measured speed of sound by the scaling factor:

$$c_{s(corr.)} = \frac{c_{s(th)}(k \to 0)}{c_{s(th)}(k_{meas.})} c_{s(meas.)}. \tag{10}$$

These values are presented in Supp. Table 2 and plotted in Fig. 4 of the main text.